# The structure and composition of high redshift radio galaxies

R. A. E. Fosbury[1]

*ST-ECF, Garching bei München, Germany, (rfosbury@eso.org)*

J. Vernet

*ESO, Garching bei München, Germany*

M. Villar-Martín

*Department of Physics & Astronomy, University of Sheffield, UK*

M. H. Cohen

*Astronomy Department, California Institute of Technology, Pasadena, USA*

A. Cimatti, S. di Serego Alighieri

*Osservatorio Astrofisico di Arcetri, Firenze, Italy*

P. J. McCarthy

*The Observatories of the Carnegie Institution, Pasadena, USA*

**Abstract.** Keck spectropolarimetry, giving spectral coverage from Ly-$\alpha$ to beyond CIII], and HST imaging of a sample of powerful radio galaxies with $z \sim 2.5$ has been obtained. These data are giving us a clear picture of the nature of the 'alignment effect' and are revealing new correlations between polarization and emission line ratios which may be interpreted in the context of the stellar evolutionary histories of these massive galaxies. In particular, we see the 2200Å dust absorption feature in the radio galaxy continua and a large variation in the NV/CIV line ratio amongst objects having a similar ionization level. VLT infrared spectroscopy of this and similar samples will give us a view of a period of galaxy history during which rapid chemical evolution was taking place.

## 1. Introduction

This is a short progress report on a rather extensive programme we are carrying out to study the structure and composition of high redshift radio galaxies

---

[1]Affiliated to the Astrophysics Division, Space Science Department, European Space Agency

(HzRG) — and, by implication, the host galaxies of radio quasars — using observations in the optical, IR and mm bands. A separation of the stellar and the AGN-related components is made using a combination of Keck spectropolarimetry and high resolution WFPC 2 imaging in the rest-frame UV (below the 4000Å break), NICMOS imaging in the rest-frame optical, and photometric measurements of cool dust thermal emission at longer wavelengths. The relevance to this meeting is the use of NICMOS to image the evolved stellar population in these galaxies (see the following talk by McCarthy) during the epoch when powerful AGN were most common. In addition, we plan to use VLT (ISAAC) IR-spectroscopy to measure the rest-frame optical emission line spectrum, allowing us to perform the kind of detailed ionization/composition analysis which has already been carried out on local objects.

## 2. The sample

We have selected RG with $z \sim 2.5$ which allow us to study the strong UV emission lines from Ly-$\alpha$ to CIII], the UV continuum, resonance absorption lines and the 2200Å dust feature in the optical band and to straddle the 4000Å break in the $1$–$2m\mu$ region.

Our principal sample constists of eight objects (six of which have already been analysed) with $2.3 < z < 2.9$ and this is suplemented by three sources from the literature having similar quality data but extending the redshift range to $1.8 < z < 3.8$. More data are currently being obtained on sources with $z > 3$.

## 3. Observations

The Keck spectropolarimetric observations for the first two sources in the programme are described in Cimatti et al. (1998). Four more sources have been observed and reduced and three further objects were observed during the period of this Workshop. An example of the spectropolarimetry is shown in Figure 1. The HST WFPC 2 and NICMOS images, where available, were taken from the public archive at the ST-ECF and from the McCarthy et al. program (ID 7498). An example of the NICMOS and WFPC 2 imaging is shown in Figure 2. Some deep groundbased imaging data have been taken from the literature to complement the higher resolution but shallower HST images.

## 4. Principal results

Here we summarise the principal results to date. These will be described more fully in papers in preparation.

- All sources show a strong 'alignment effect' between their UV and radio morphologies although the structures are complex. One case, 4C 23.56 (Knopp & Chambers 1997), shows a beautiful 'ionization cone' in Ly-$\alpha$. The brightest UV emission is extended and does not necessarily coincide with the nucleus (radio core).

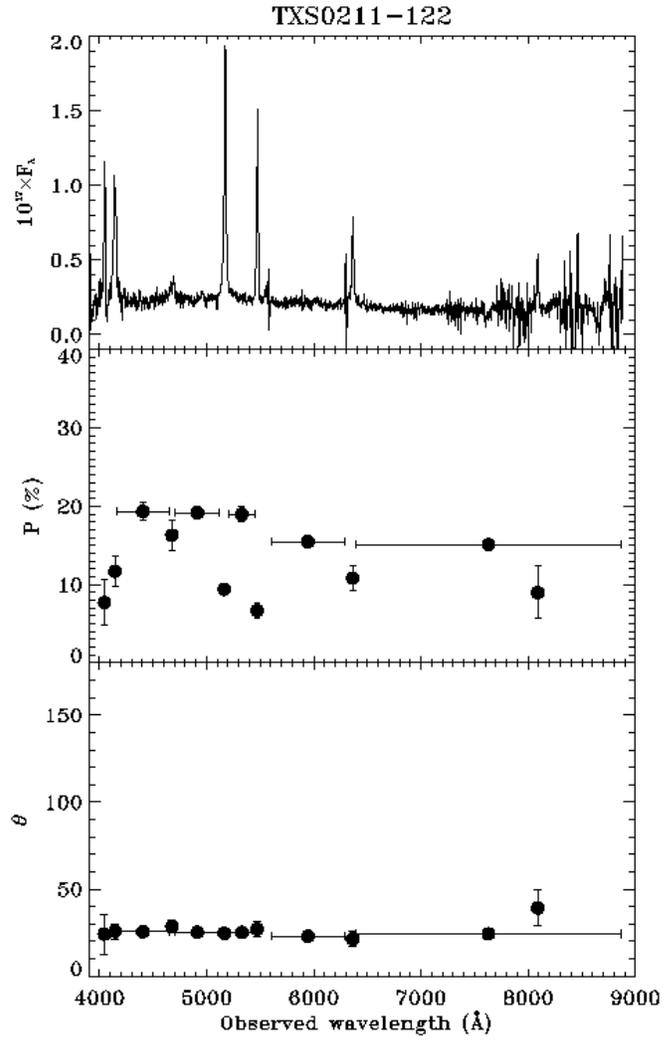

Figure 1. A Keck II, LRISp spectropolarimetric observation of the radio galaxy TXS 0211-122. The three panels show respectively the total flux (in $10^{-17}$ erg cm$^{-2}$ s$^{-1}$ Å$^{-1}$), the fractional polarization in continuum (wide horizontal bars) and line bands and the position angle of the $E$-vector. The strong emission lines are, from short wavelengths, Ly-$\alpha$, NV, CIV, HeII and CIII].

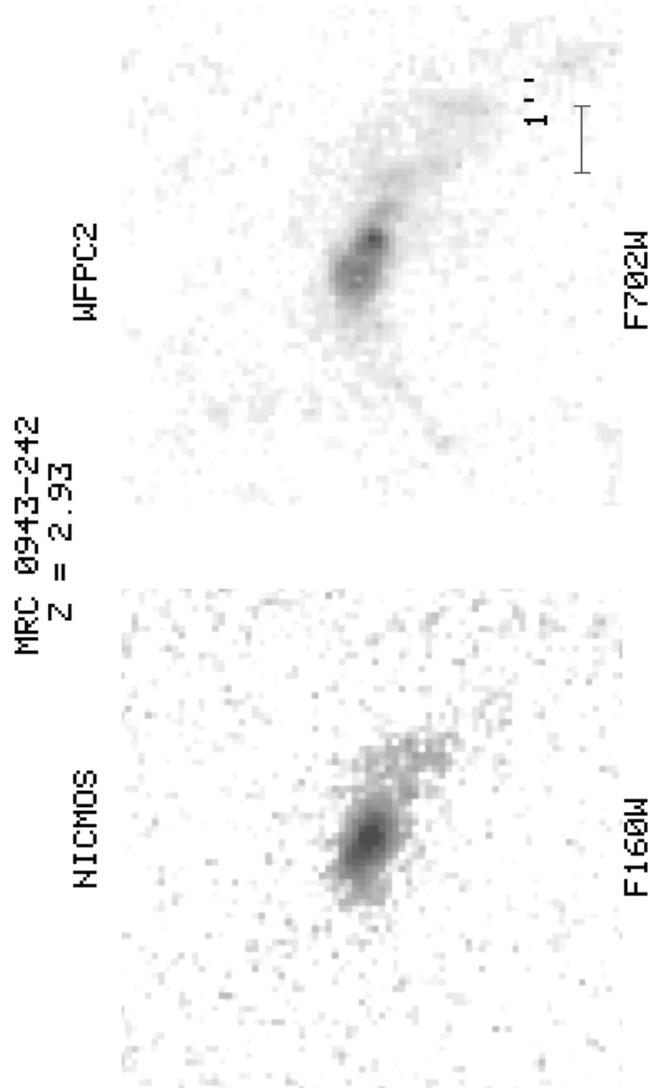

Figure 2. Images from NICMOS/F160W and WFPC 2/F702W of the $z = 2.93$ radio galaxy MRC 0943-242 represented at the same scale. This object has an aligned component which still dominates in the H-band. Although bluer than the underlying galaxy, the aligned light is somewhat redder than usual in these objects although the range in UV colours is small (see text). There is a prominent Ly-$\alpha$ absorption component.

- The continuum colours are remarkably similar to one another and can be fitted by a power law absorbed by a standard Galactic (in the RG restframe) extinction law with $E_{B-V} \sim 0.1$, which corresponds to $\tau \sim 1$ at 1500Å. Example fits to three of the sources are shown in Figure 3.

- Interstellar absorption lines are seen and, in some objects, there is evidence for wind and photospheric absorption lines from hot stars. Several sources show complex, spatially extended absorption structures at Ly-$\alpha$ (see also van Ojik et al., 1997).

- The emission lines are spatially extended (up to $\sim$ 20 arcsec or $\sim$ 150 kpc for Ly-$\alpha$) and show complex kinematic structures extending over $\pm 2000$ km s$^{-1}$.

- The continuum linear polarization, measured just longward of Ly-$\alpha$/NV, ranges from $< 3\%$ to $\sim 20\%$. The $E$-vector is perpendicular to the UV extension as seen at HST resolution (but not necessarily precisely to the radio axis).

- The emission line spectra indicate a rather constant level of ionization with a small range in the observed CIII]/CIV and HeII/CIV line ratios.

- Amongst the spatially integrated properties, the strongest correlations, shown in Figure 4, are observed to be:

    1. between continuum polarization, $P$, and the Ly-$\alpha$/CIV emission line ratio (anticorrelation)
    2. between $P$ and the NV/CIV ratio

## 5. Discussion

The conclusions we are drawing from these studies can be summarised as follows:

- Powerful radio galaxies contain (hidden) QSO nuclei whose EUV emission ionizes the extended gas along the radio axis and whose FUV emission we see scattered by extended dust structures.

- The scattered component can — but does not always — dominate the observed UV continuum and there is evidence for an unpolarized hot stellar component in addition to a nebular continuum contribution from the emission line gas. The presence of an old, red stellar population becomes apparent above the 4000Å break, although this does not always dominate the observed flux in the infrared bands.

- There is a direct connection between the scattering mechanism (which produces the polarization) and the destruction of Ly-$\alpha$ photons. This could simply result from the abundance and spatial distribution of dust, although orientation effects may be important as well.

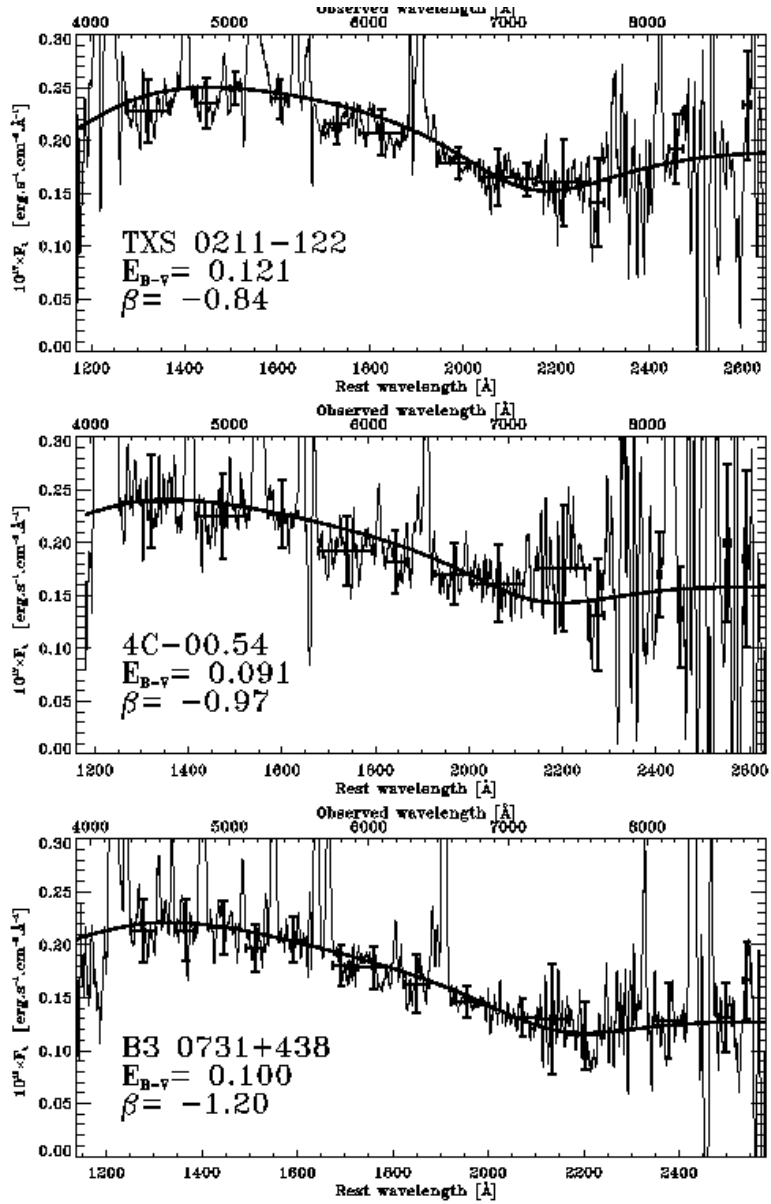

Figure 3. Total flux spectra of three of the radio galaxies scaled to show the continuum. The crosses mark continuum bins chosen to be free of emission lines and atmospheric absorption features with the vertical bars representing one sigma statistical errors. The fitted curves are two parameter fits of a power law (with index $\beta$ in $F_\lambda$) absorbed by a Galactic extinction curve in the rest-frame of the RG. The derived values from $1/\sigma^2$ weighted fits are shown in the labels.

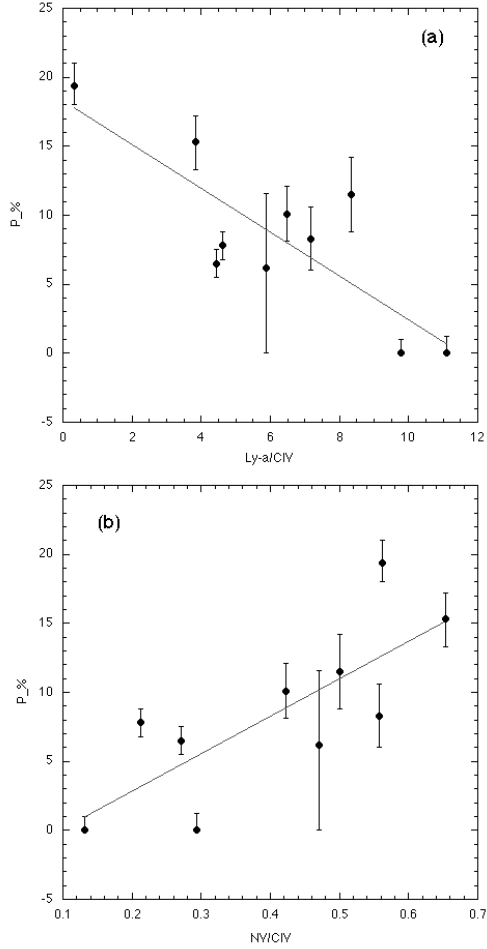

Figure 4. Plots of the fractional continuum polarization against (a) the Ly-$\alpha$/CIV emission line ratio and, (b) the NV/CIV ratio. The sources in the plots are: 4C 23.56 (components a and b, $z = 2.482$), 4C -00.54 ($z = 2.366$), TXS 0211-122 ($z = 2.338$), B3 0731+438 ($z = 2.429$), USS 0828+193 ($z = 2.572$) and MRC 0943-242 ($z = 2.93$) from our own observations and 4C 41.17 ($z = 3.798$), MRC 2025-218 ($z = 2.63$) and 3C 256 ($z = 1.824$) from the literature.

- The continuum exhibits dust extinction signatures in the form of the 2200Å dip and a peak at the position of the extinction minumum around 1400Å. Some of the extinction may arise in an extended halo outside the regions which see the QSO radiation field directly. This would be consistent with the absence of the 2000Å dip in radio quasars.

- There appears to be strong connection between the dominance of scattered light (dust abundance and/or intrinsic quasar luminosity) and nitrogen/carbon ratio. The behaviour of the NV/CIV, NV/HeII diagram indicates that the effect is due to nitrogen abundance variations and not to carbon depletion. This may be related to the suggestion of a relative overabundance of nitrogen in high redshift QSOs (Hamann & Ferland 1993).

These objects are telling us the story of the formation of massive galaxies and their quasar nuclei during the epoch when AGN had their maximum space density. The UV emission lines can give us some clues to the chemical composition of the extended nebulosity but to make inferences with more confidence, we need to measure the optical forbidden line spectrum in the infrared with ISAAC at the VLT.

**Acknowledgments.** We thank Laura Pentericci for making available to us the reductions of the NICMOS images. We are grateful to Bob Goodrich for frequent help with the polarimetric observations and many discussions. Our Keck programme is supported by NATO Collaborative Research Grant number 971115. This paper is based partially on observations made with the NASA/ESA Hubble Space Telescope, obtained at the Space Telescope Science Institute, which is operated by the Association of Universities for Research in Astronomy, Inc., under NASA contract NAS 5-26555.